\documentclass[fleqn,usenatbib]{mnras}

\usepackage{newtxtext,newtxmath}

\usepackage[T1]{fontenc}

\usepackage{longtable}
\usepackage{booktabs, multirow} 
\usepackage{soul}
\usepackage[table]{xcolor} 
\usepackage{changepage,threeparttable} 

\DeclareRobustCommand{\VAN}[3]{#2}
\let\VANthebibliography\thebibliography
\def\thebibliography{\DeclareRobustCommand{\VAN}[3]{##3}\VANthebibliography}

\usepackage{graphicx}
\setkeys{Gin}{width=\columnwidth}
\usepackage{amsmath}	
\usepackage{multirow}
\usepackage{supertabular}
\newcommand{\msun}{\text{M}_{\odot}}

\title[SLSN host galaxies]{Metallicity beats sSFR: The connection between superluminous supernova host galaxy environments and the importance of metallicity for their production}

\author[C. Cleland \& S. L. McGee \& M. Nicholl]{
Cressida Cleland,$^{1}$\thanks{E-mail: cressidac@star.sr.bham.ac.uk}
Sean L. McGee,$^{1}$
Matt Nicholl$^{2}$
\\
$^{1}$School of Physics and Astronomy, University of Birmingham, Edbgaston, Birmingham B15 2TT, UK \\
$^{2}$Astrophysics Research Centre, School of Mathematics and Physics,
Queen’s University Belfast, Belfast BT7 1NN, UK
}

\date{Accepted XXX. Received YYY; in original form ZZZ}

\pubyear{2022}

\begin{document}
\label{firstpage}
\pagerange{\pageref{firstpage}--\pageref{lastpage}}
\maketitle

\begin{abstract}
We analyse 33 Type I superluminous supernovae (SLSNe) taken from ZTF's Bright Transient Survey to investigate the local environments of their host galaxies. We use a spectroscopic sample of galaxies from the SDSS to determine the large-scale environmental density of the host galaxy. Noting that SLSNe are generally found in galaxies with low stellar masses, high star formation rates, and low metallicities, we find that SLSN hosts are also rarely found within high-density environments. Only $3\substack{+9 \\-1}$ per cent of SLSN hosts were found in regions with 2 or more bright galaxies within 2 Mpc. For comparison, we generate a sample of 662 SDSS galaxies matched to the photometric properties of the SLSN hosts. This sample is also rarely found within high-density environments, suggesting that galaxies with properties required for SLSN production favour more isolated environments. Furthermore, we select galaxies within the Illustris-TNG simulation to match SLSN host galaxy properties in colour and stellar mass. We find that the fraction of simulated galaxies in high-density environments quantitatively matches the observed SLSN hosts only if we restrict to simulated galaxies with metallicity  $12+\log($O/H$) \leq 8.12$. In contrast, limiting to only the highest sSFR galaxies in the sample leads to an overabundance of SLSN hosts in high-density environments. Thus, our measurement of the environmental density of SLSN host galaxies appears to break the degeneracy between low-metallicity or high-sSFR as the driver for SLSN hosts and provides evidence that the most constraining factor on SLSN production is low-metallicity. 
\end{abstract}




\section{Introduction}

While the detection rates of supernovae have rapidly increased over the last decade or so, thanks to wide-field high-cadence surveys such as Pan-STARRs \citep{chambers2016}, ATLAS \citep{tonry2018} and ZTF \citep{bellm2019}, superluminous supernovae (SLSNe) remain a rare and elusive transient event. SLSNe were originally classed as a supernova event that is 10-100 times brighter than typical supernovae \citep{quimby2011,galyam2012}, however are now classified by their unique spectra \citep[e.g.][]{Lunnan2018,Quimby2018,Angus2019}. Their lightcurves differ from those of other supernovae as they are both broader and brighter. There are two classes of SLSNe: Type I and Type II. Type I SLSNe (often referred to as simply SLSNe, a convention we adopt in this work) are dominated by OII absorption lines at maximum luminosity, while Type II SLSNe exhibit sharply peaked hydrogen emission lines from circumstellar interaction. The physical mechanisms which cause Type I SLSNe remain a topic of debate, with mechanisms such as a central engine in the form of a magnetar \citep[see e.g.][]{kasen2010,inserra2013,Nicholl2017} or interaction with the circumstellar medium \citep[see e.g.][]{smithmccray2007,chevalier2011,chatzopoulos2012,ginzburgbalberg2012} being proposed as possible power sources for the increased luminosity. 

\begin{table*}
\caption{Descriptions of each of the samples used. The ranges shown in the redshift and magnitude columns refer to the minima and maxima of that property for each sample, not necessarily the selection constraints.}
\centering

\begin{tabular*}{\textwidth}{c @{\extracolsep{\fill}} cccccc}
    Sample & \# obj. & Redshift & Median(z) & Magnitude & Median($M_i$) & Description \\ \hline\hline
    
    SLSN-g & 33 & $0.064\leq z\leq 0.39$ & 0.159 & $-22.74 \leq M_i\leq-15.32$ & -18.41 & SLSN host galaxies \\ \hline
    
    SDSS-m & 662 & $0.036\leq z\leq 0.312$ & 0.117 & $-22.196 \leq M_i\leq -16.709$ & -19.632 & SDSS photometric-z galaxies matched to SLSN-g \\\hline
     
    SN-Ia & 3268 & $0.001\leq z\leq 0.0.156$ & 0.065 & $-26.56\leq M_i\leq -11.56$ & -20.76 & Type Ia supernova host galaxies\\ \hline
    
    Spec-z & 300000 & $0.0\leq z\leq 0.4$ & 0.144 & $-31.214\leq M_i\leq -10.006$ & -22.385 & SDSS spectroscopic-z galaxies \\
    \hline
  \end{tabular*}
  \label{tab:table}
\end{table*}

We can use properties of their host galaxies to infer the conditions required for a SLSN to occur. This can help constrain the processes at work. SLSNe have been found mostly in low-mass galaxies, with very few being found in host galaxies with stellar mass $>10^8$ M$_\odot$ \citep{schulze2018}. This is surprising, as one might expect that with more stars comes more supernova events (\cite{Sullivan2006,Smith2012,Wiseman2021}, see also \cite{Perley2016}). Various authors have found that SLSN host galaxies typically have high specific star-formation rates \citep[$\sim 10^{-9}$ yr$^{-1}$,][]{neill2011,leloudas2015,angus2016}, leading to the idea that SLSNe may be the first stars to explode following a starburst. {\citet{leloudas2015} note that SLSN host galaxies have properties which are consistent with extreme emission line galaxies (EELGs), i.e. high-sSFR and also highly ionised gas. }
Meanwhile, \citet{lunnan2014}, \citet{Chen2017} and \citet{perley2016host} argue that the low-metallicity observed in SLSN hosts is the more important factor in the production of these rare transients, {and that their low metallicities can explain the extreme energies seen \citep{chen2013}}. Nevertheless, since dwarf galaxies tend to have both low metallicities and bursty star-formation histories, {and the production of SLSNe appears averse to high-metallicity and low specific star-formation rate \citep{schulze2018}},  there is a degeneracy regarding which of these factors is more critical.

\begin{figure*}
    \centering
    \includegraphics[width=\textwidth]{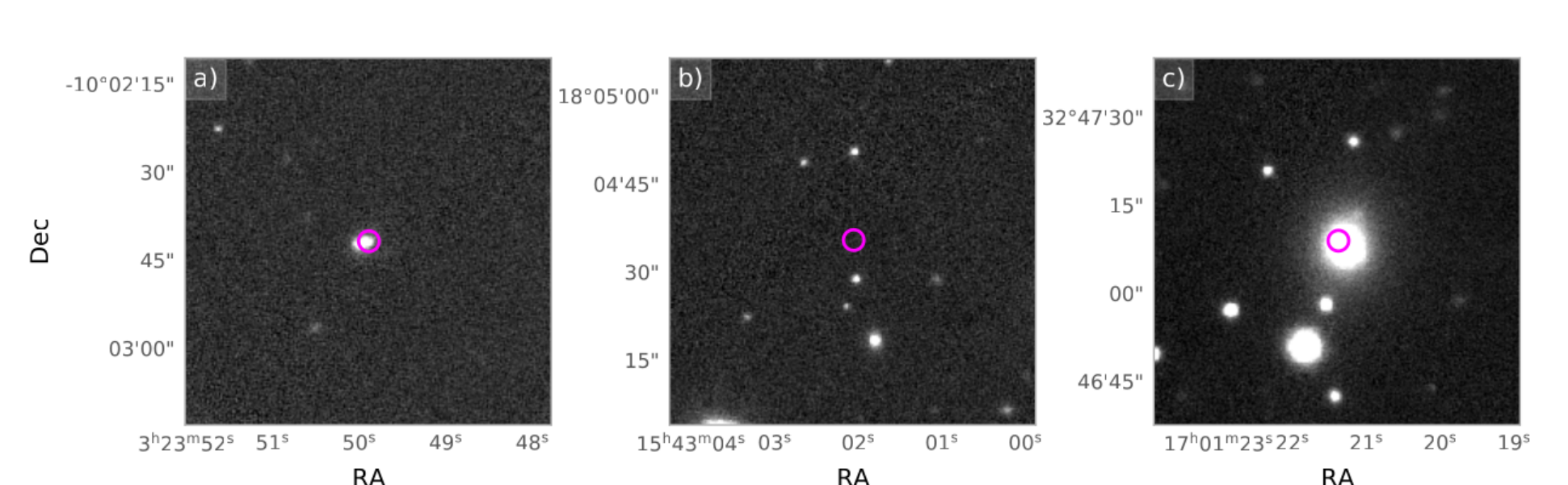}
    \caption{PanSTARRS $i$-band cut-out images \citep{flewelling2020} of SLSN sources and of a SN Type Ia host. Panel a) shows object ZTF21aaarmti, panel b) shows object ZTF22aalzjdc and panel c) shows object ZTF18aangpkx. The magenta circle in each panel depicts the location of the source. In panel b), no host galaxy was found for the source. The host galaxy in panel c) is in a higher-density environment than the SLSN hosts; see Section \ref{sec:obs} for details. The absolute magnitude of the host galaxy in panel a) is $M_i=-21.36$ and its redshift is $z=0.193$. The absolute magnitude of the host galaxy in panel c) is $M_i=-22.67$ and its redshift is $z=0.070$. The redshift of the SLSN source in panel b) is $z=0.200$. The size of each of the images is $62.50''$.}
    \label{fig:ps}
\end{figure*}

It is well known that the local environment of galaxies, that is whether a galaxy is located within a galaxy group or a galaxy cluster or it is isolated, has a profound effect on a number of galaxy properties, including the metallicity \citep{tremonti2004,ellison2009, cooper2009}, star formation rate \citep{wetzel2013,cleland2021} and overall evolution of the galaxy \citep{dressler1980}. Gravitational interactions between galaxies can disturb the gas and dust which fuels star-formation, leading to starburst phases, or similarly galaxies may merge which will also have an effect on the star-formation rates of the galaxies involved. Conversely, high-density environments may cause the quenching of star-formation in an infalling galaxy. It naturally follows to investigate the local environment of SLSNe host galaxies to uncover any environmental factors that may impact SLSNe rates. \citet{orum2020} find that up to 50 per cent of superluminous supernova host galaxies have at least one companion within 5 kpc, which they use to explain the increased star-formation rates required for SLSNe to occur. However, for the purposes of this work we are interested in the overall density of the local environment of the SLSN host galaxy. That is, a close companion galaxy of arbitrary stellar mass within kiloparsecs of the supernova host galaxy does not necessarily lead to the same environmental effects at larger scales (e.g., Mpc). In order to probe for such effects, we aim to use bright galaxies as a proxy for density, noting that this is a first order approach, with the absolute magnitude of galaxies roughly tracing stellar mass. By investigating the environments of these galaxies in this way, we aim to break the degeneracy between the requirements of low-metallicity and high specific star-formation rate for SLSN production.

In this work we identify isolated SLSNe and SLSNe in groups, and compare between various host galaxy properties. We also compare against Type Ia supernova host galaxies, which occur in a wider variety of host galaxies. We then investigate these host galaxies further within the Illustris-TNG simulation suite, by exploring the metallicities and specific star-formation rates of a sample of simulated galaxies with observational properties similar to those of SLSN hosts.

The paper is structured as follows: in Section \ref{sec:data}, we describe the data used and how it was obtained; in Section \ref{sec:res} we explain the results and discuss their implications; finally in Section \ref{sec:conc} we summarise our findings. For the computation of cosmological distances, we use a flat universe with $H_0 = 70.2$ and $\Omega_m=0.277$.

\begin{figure*}
    \centering
    \includegraphics[width=\textwidth]{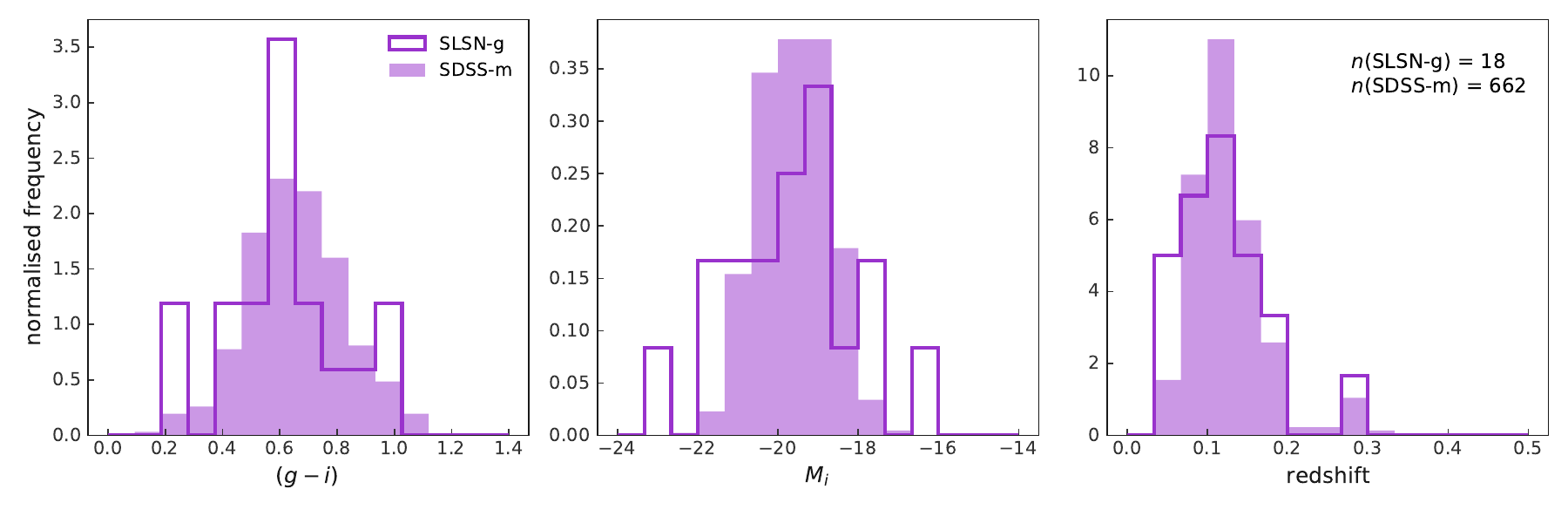}
    \caption{Histograms of the colours, absolute magnitudes, and redshifts of galaxies in SLSN-g with reliable photometry compared to the matched SDSS sample. {KS-tests result in p-values of 0.46, 0.36, and 0.75 respectively, validating that the SLSN-g and SDSS-m samples match each other.}}
    \label{fig:comparison}
\end{figure*}

\section{Data}\label{sec:data}

We obtain coordinates, redshifts and host photometry ($g-i$ colours and $M_i$ absolute magnitudes) of 55 Type I SLSNe (hereafter simply referred to as SLSNe) and 4816 Type Ia supernovae from the ZTF Bright Transient Survey\footnote{\url{https://sites.astro.caltech.edu/ztf/bts/bts.php}} \citep[BTS;][]{masci2019,perley2020}. This was the full sample of Type I SLSNe and Type Ia SNe available on the BTS at the time of analysis (September 2022). The BTS is used to obtain properties on supernova events and their host galaxies due it being a well-defined sample with a limiting magnitude of $\approx 18.5$ mag that selects for nearby events with a higher chance of host detection. The supernovae in this sample have been classified spectroscopically, and by a variety of research groups. Table \ref{tab: slsnelist} lists and cites the classification group for each SLSN. The code \textsc{Sherlock}\footnote{\url{https://github.com/thespacedoctor/sherlock}} is used to cross-identify for a host galaxy and its photometry \citep{smith2020}; this information is available directly on the \textsc{Lasair} broker \citep{lasair} for each source. Some sources have no catalogued host, and are marked as `orphans'. See Figure \ref{fig:ps} for examples of a SLSN source with a host detection (left panel) and a SLSN source where no host was detected (middle panel). We denote the SLSN host galaxy sample as SLSN-g, and the Type Ia supernova host galaxy sample as SN-Ia. 

We obtain a spectroscopically-redshifted sample of $300000$ random galaxies (Spec-z) from the Sloan Digital Sky Survey (SDSS) DR17 \citep{smee2013,blanton2017,abdurrouf2022}, such that their redshifts are local and precisely-known, $0<z<0.4$. We also obtain 16494 random photometrically-redshifted galaxies from SDSS DR17 \citep{fukugita1996,gunn1998,gunn2006,doi2010} to better match the parameter space of the SLSNe sample, with $0<z<0.4$, such that $z_\text{err}<0.1$ and $z_\text{err}/z<0.5$. Both samples were obtained using the SDSS query service CasJobs\footnote{\url{http://cas.sdss.org/dr17/}}. For each sample of supernova hosts, we restrict to sources within the on-sky footprint of the spectroscopic sample, resulting in 35 SLSNe (18 with detected hosts) and 3450 Type Ia supernovae.

\subsection{Matched sample generation}

Since the number of SLSN host galaxies is so low, it is desirable to generate a matched galaxy sample, based on SLSN host galaxy properties. This means we can make statistical comparisons of SLSN host-like galaxies, but on a much larger sample of galaxies. The properties of the SLSN host galaxies are based on their photometry, so we generate a sample of galaxies that are matched based on photometric properties such as absolute magnitude and colour. We generate this matched sample to the SLSN host galaxies from the SDSS photometric galaxies \citep{cooper2009}. For each SLSN host galaxy, we randomly search for a photometric galaxy within a sphere in parameter space according to Equation \ref{eqn:match}:
\begin{equation}\label{eqn:match}
    \left(\frac{\Delta\text{colour}}{0.2} \right)^2 + \left(\frac{\Delta M_i}{1.2}\right)^2 + \left(\frac{\Delta z}{0.04}\right)^2 =1,    
\end{equation}
where $\Delta$colour, $\Delta M_i$ and $\Delta z$ refer to the difference between the $(g-i)$ colour, absolute magnitude, and redshift of the SLSN host galaxy and the photometric galaxy. The scaling factor in the denominator of each term in Equation \ref{eqn:match} comes from taking the range of each property and multiplying by a factor of {0.1. This factor was chosen by visual inspection to ensure an adequate chance of finding an appropriate match, while reducing mismatches in parameter space.} Applying this factor ensures each property has the same weight in parameter space. {We attempt a random search for a unique SDSS galaxy that satisfies Equation \ref{eqn:match} on a SLSN host galaxy; this search gets repeated until a match is found, to a maximum of 1000 times. This process is repeated on SLSN host galaxies 10000 times. This results in a matched sample (SDSS-m) of 662 galaxies. Note that varying the multiplicative factor in Equation \ref{eqn:match}, and the number of repetitions performed on each galaxy, has no quantitative effect on the results discussed in Section \ref{sec:obs}.} For this matching procedure, we only use the 18 SLSN sources with detected hosts and reliable host photometry. The rest of the analysis depends on the positions and redshifts of each SLSN source directly, and so the entire sample of 33 SLSNe is used. 

The details of each sample are listed in Table \ref{tab:table}. Histograms of the distributions of the colour, absolute magnitude, and redshift of SLSN-g and SDSS-m are plotted in Figure \ref{fig:comparison}. {Kolmogorov-Smirnov tests validate that the distributions match well}, with the added benefit of {over an order of magnitude more galaxies in the matched sample than SLSN host galaxies.}

\section{Results and discussion}\label{sec:res}
\subsection{Observations}\label{sec:obs}

\begin{figure}
    \centering
    \includegraphics{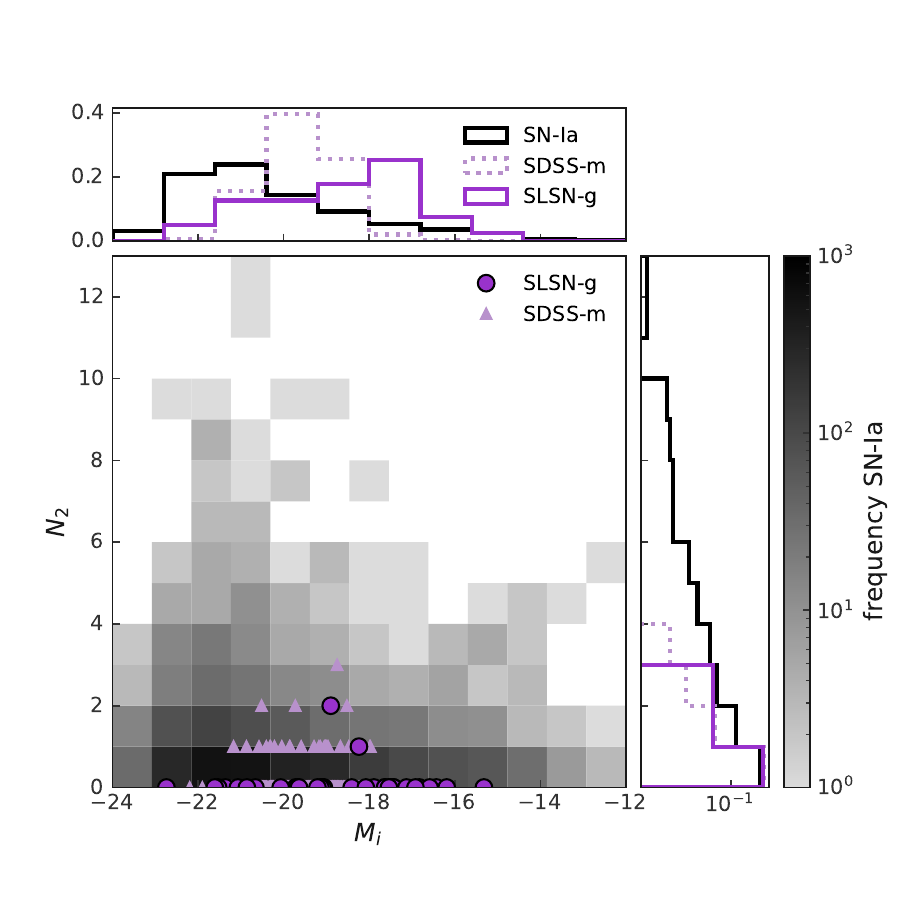}
    \caption{A scatter plot of the number of bright Spec-z galaxies within 2 Mpc of a target galaxy against absolute magnitude, compared to a 2d histogram of the same distribution for Type Ia SNe host galaxies. SLSN-g is shown as dark purple circles and the matched sample, SDSS-m, is shown as light purple triangles. Normalised histograms are included on the $x-$ and $y-$axes and show the absolute magnitude and $N_2$ distributions for SN-Ia, SDSS-m and SDSS-g.}
    \label{fig:count}
\end{figure}

Using bright, well-observed galaxies as a proxy for density, we count the number of bright ($M_i<-21.5$) Spec-z galaxies that are at an on-sky projection of 2 Mpc or less of the target galaxy, within $\Delta z = 0.005$, {based off a typical velocity dispersion of a galaxy group/cluser \citep{yang2007}}. We denote this quantity as $N_2$. We use $M_i<-21.5$ as the threshold for bright galaxies as this {magnitude is the completion limit for the entire spectroscopic sample for this redshift range}, and we use 2 Mpc as it is the upper end of galaxy group diameters \citep{yang2007}. Note that varying the $M_i$ threshold does not have a qualitative effect on the results. We calculate $N_2$ for SLSN-g, SDSS-m, and SN-Ia as a control. SN-Ia works as a control sample because these galaxies are {much less confined to a specific region of parameter-space compared to SLSNe}, and the photometry is easily comparable to SLSN-g since both samples come from ZTF. A 2-d histogram showing these results is shown in Figure \ref{fig:count}, with $N_2$ shown as a function of absolute magnitude. Note the log scale in the $z$-axis. This quantity is also plotted for SLSN-g and SDSS-m in dark purple and light purple, respectively. We can see that while the distribution $N_2$ for SN-Ia occupies a large and varied region of parameter-space (note 70 per cent of SN-Ia still has $N_2=0$), $N_2 \leq 2$ for SLSN-g and $N_2 \leq 3$ for SDSS-m, with the majority of galaxies having $N_2=0$. This is not simply a consequence of reduced numbers of SLSN-g compared to SN-Ia; binned by magnitude, the fraction of SLSN-g with any bright neighbour (i.e. $N_2\geq1$) reaches a maximum of about 10 per cent\footnote{This number is entirely dependent on the choice of binning since $N_2 \approx 1$ for SLSN-g in any magnitude bin}, which is only half of the minimum of the fraction of SN-Ia with any neighbour in the same bins. Similarly, SDSS-m only reaches a maximum of about 10 per cent. 

\begin{figure}
    \centering
    \includegraphics[width=\columnwidth]{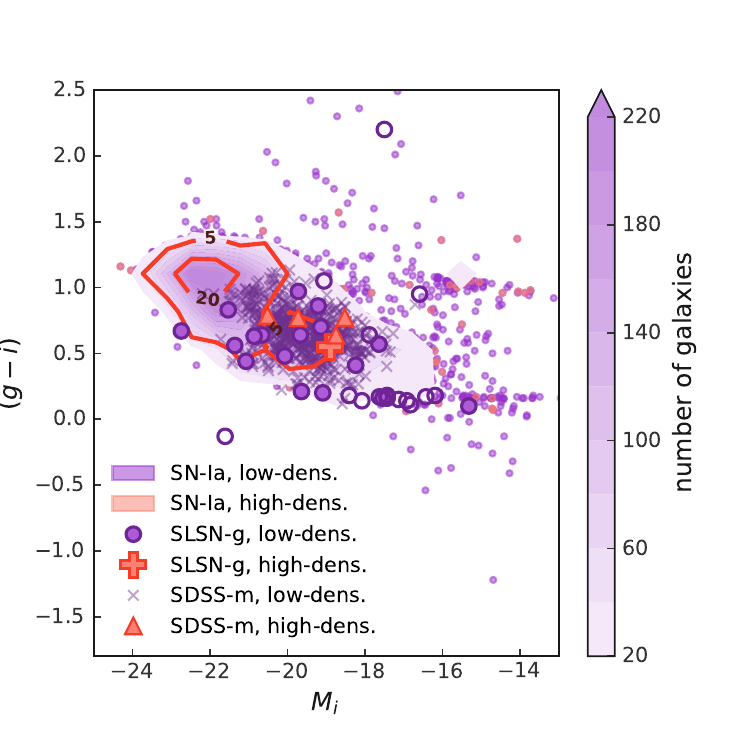}
    \caption{A colour-magnitude diagram comparing SLSN-g and SDSS-m to SN-Ia. Low-density SLSN-g galaxies are plotted in purple circles, high-density SLSN-g galaxies are plotted in orange pluses, low-density SDSS-m galaxies are plotted in purple crosses, high-density SDSS-m are plotted in orange triangles, and the distributions for low- and high-density SN-Ia galaxies are plotted as purple filled in contours and orange contour lines respectively. Open purple circles represent the SLSN-g sources with unreliable host photometry. Purple and orange points shown individual SN-Ia galaxies when they are not captured by the contours.}
    \label{fig:cmd}
\end{figure}

We consider galaxies with $N_2 \geq 2$ as being in a high-density environment, with everything else being in a low-density environment. This number is chosen under the assumption that a typical {galaxy group or cluster halo is of the order of a few Mpc in diameter \citep{yang2007}, so any more than two bright (large) galaxies would be considered a high-density environment. We emphasise the fact that this is a generous constraint, set in an attempt to maximise the number of SLSN hosts in high-density environments. An example of a SN-Ia host galaxy with a high $N_2$ number ($N_2=8$) is shown in the right panel of Figure \ref{fig:ps}}. We plot SLSN-g, SDSS-m, and SN-Ia on a colour-magnitude diagram in Figure \ref{fig:cmd}, separated by high- and low-density. Contour lines map out the SN-Ia galaxies that are in high-density environments. We see that the high-density SLSN-g and SDSS-m galaxies are found on the outer regions of these contour lines. That is, they preferentially avoid the region in phase-space where high-density environments occur. This result, along with finding that the matched sample galaxies are also almost exclusively found in low-density environments, is indicative that the galactic properties of SLSN hosts are the constraining factor in the density of the environment of these galaxies. In other words, the type of galaxies where SLSNe are found (blue, low-mass, low-metallicity) are rarely found in high-density environments. {As a fraction of each sample, SLSNe that occur in high-density environments is $1/33 = 0.03\pm^{0.06}_{0.01}$, and matched galaxies that occur in high-density environments is even lower, at $4/662 = 0.006\pm^{0.005}_{0.002}$.}

\begin{figure*}
    \centering
    \includegraphics[width=\textwidth]{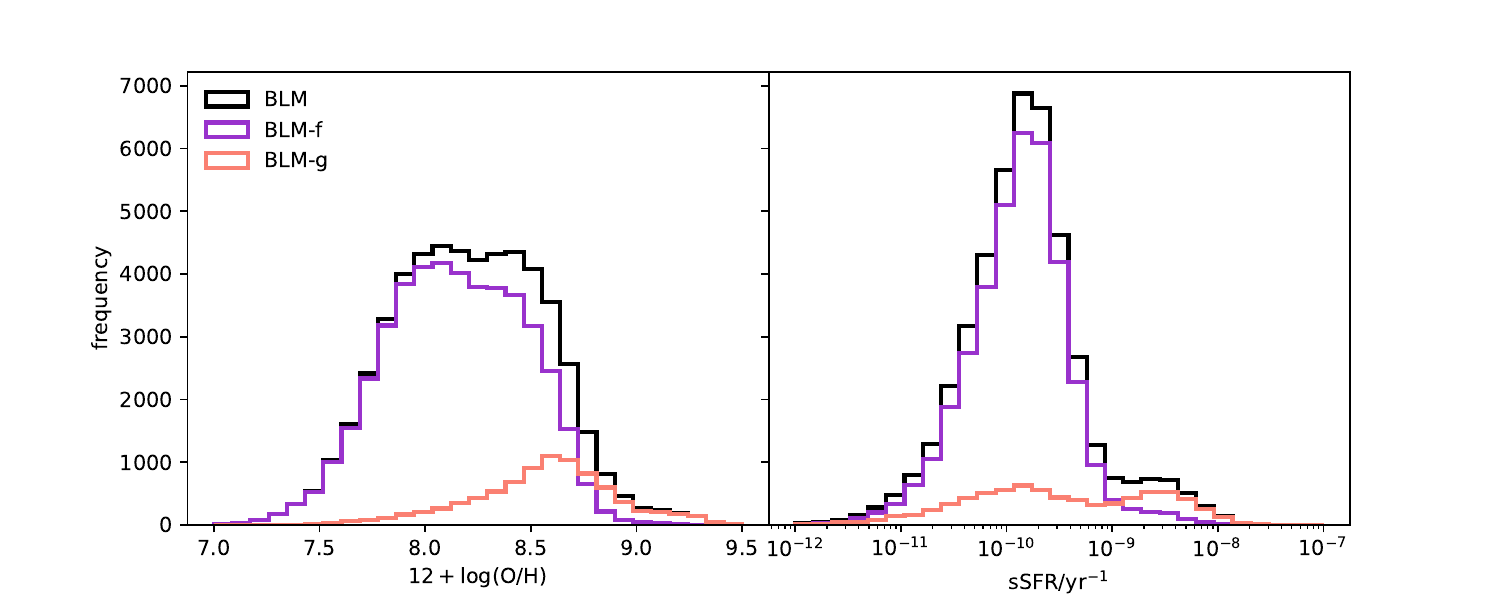}
    \caption{Histograms of metallicity (left) and sSFR (right) in IllustrisTNG for all blue low-mass (BLM) galaxies (black), BLM field galaxies (purple) and BLM group galaxies (orange).}
    \label{fig:metallicity-ssfr}
\end{figure*}

\subsection{Simulations}
Since these galaxies are so faint, often the reliable host photometry or spectroscopy required to accurately measure physical properties like metallicity and star-formation rate is not available. Therefore it can be useful to use simulations to distinguish between real physical properties, and observational constraints. In Section \ref{subsec:sims} we utilize the IllustrisTNG suite of simulations\footnote{\url{https://www.tng-project.org/data/}}
\citep{pillepich2018,springel2018,nelson2018,naiman2018,marinacci2018,nelson2019}. IllustrisTNG is a set of large-scale cosmological, gravo-magnetohydrodynamical simulations based on the \textsc{Arepo} code \citep{springel2010}. We use the TNG-100 simulation, which has a volume of $106.5^3$ cMpc. This simulation provides the best balance of large haloes {(i.e. $M_h>10^{12}\ \msun$) and smaller resolution subhaloes (i.e. $M_h<10^9\ \msun$), although qualitatively similar results were found with the use of TNG-300.} From this simulation, we use stellar masses, star-formation rates, gas-phase metallicity abundances, galaxy colours, group halo masses, and information about group membership.

\subsection{IllustrisTNG}\label{subsec:sims}

To investigate the {tendency of SLSN host-like galaxies to avoid high-density environments further}, we make use of the IllustrisTNG suite of simulations. From the TNG100-1 simulation, we obtain stellar masses, gas metallicities, star-formation rates, and $g$ and $i$ photometry on all 4371211 subhaloes in the most recent redshift snapshot. Each subhalo resides in a parent halo with a total halo mass $M_{200}$, defined as the total mass of a sphere whose mean density is 200 times the critical density of the Universe. In line with the properties of SLSN host galaxies, we define a sample of blue, low-mass galaxies (BLM), with $(g-i)<1$ and $10^7<M_*/\msun<10^9$. {This stellar mass range has an upper limit in line with observations of SLSN host galaxies \citep[e.g. ][]{schulze2018}, and a lower limit to avoid contamination from haloes not yet formed into galaxies. Note that at this stellar mass range, virtually all galaxies ($\sim 98$ per cent) have $(g-i)<1$ and also are star-forming.} For each subhalo, we retrieve the number of subhaloes ($N_\text{subs}$) in its parent halo. For a galaxy to be in a group, we require $N_\text{subs}>1$ and $M_{200}>10^{12}\ \msun$. {This is to ensure we avoid including field galaxies with smaller (non-galaxy) haloes associated with them.} Everything else is assigned as being in the field. 

In Figure \ref{fig:metallicity-ssfr}, we plot a histogram of the log of the metallicity of BLM and a histogram of the log of sSFR, separated by group membership. Immediately it is clear that this sample is biased towards low metallicities, as is expected from observations of SLSN host galaxies (see Figure 11 in \citet{chen2015}, and Figure 8 in \citet{perley2016host}). Notably there are much fewer BLM galaxies in groups than in the field, with group galaxies making up 35.5 per cent of the BLM sample. Additionally, the BLM galaxies that reside in groups lie at the high end of the metallicity distribution. We also see that BLM galaxies typically have intermediate sSFR, as is seen in the properties of SLSN host galaxies. However, BLM galaxies in groups dominate over field galaxies at high-sSFR. 

Thus, in agreement with our observational results in the previous section, the simulation results qualitatively suggest that the reason SLSN-g are not found in dense environments is because the galaxies that host SLSN, blue low-mass, are not often found in groups. However, quantitatively, there is a potential discrepancy. From the simulation results in Figure \ref{fig:metallicity-ssfr} we measure that 35.5 per cent of such BLM galaxies are found in groups within the simulation. This means, observationally, we would expect about 12(235) of SLSN-g(SDSS-m) to be in high-density environments, compared to the 1(4) in SLSN-g(SDSS-m). 

\begin{figure}
    \centering
    \includegraphics{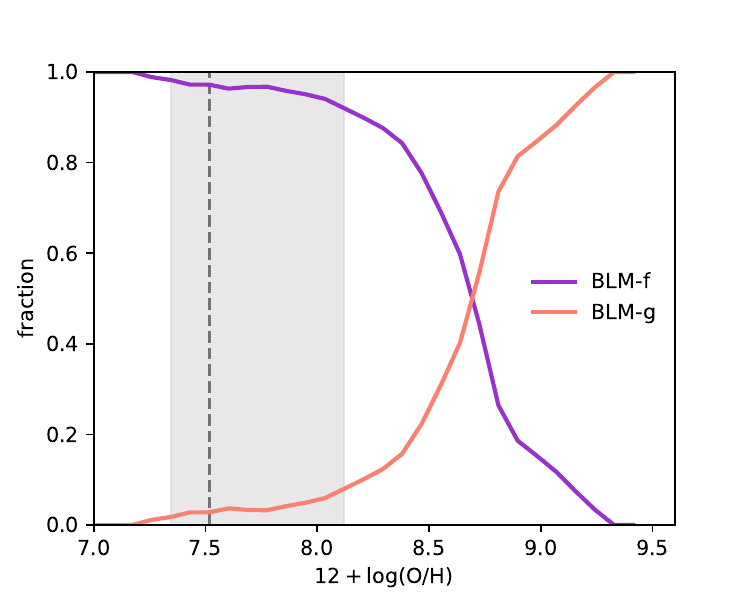}
    \caption{Fractions of BLM-f (purple) and BLM-g (orange) of the total BLM sample with respect to metallicity. The gray dashed line shows the metallicity at which group galaxies account for 3 per cent of the entire BLM sample, in accordance with observations, with upper and lower limits shaded in gray.}
    \label{fig:fraction}
\end{figure}

This potential discrepancy is pointing to an additional requirement in our simulated sample to properly recover the observed environmental distribution -- that of low-metallicity. It is known that SLSNe require host galaxies with low metallicities \citep[e.g.][]{lunnan2014}, and in the simulation results of Figure \ref{fig:metallicity-ssfr}, the lower metallicity ranges are dominated by field galaxies. We better illustrate this in Figure \ref{fig:fraction}, where we plot the fractions of field galaxies and group galaxies with respect to the total BLM sample as a function of metallicity. Field galaxies clearly dominate at $Z< Z_\odot$, and group galaxies dominate at $Z>Z_\odot$; here we assume $Z_\odot=8.69$ following \citet{asplund2009}. We also plot the metallicity at which the fraction of observed SLSN host galaxies in groups is 0.03. We can then apply generous constraints by considering the uncertainties on the SLSN occurrence rate above. By computing beta distribution confidence intervals at $c=0.68$ \citep{cameron2011} on the fraction of observed SLSNe in groups we find this fraction to be $0.03\substack{+0.09 \\-0.02}$. In Figure \ref{fig:fraction} these confidence intervals are shaded in gray. Even with such a wide confidence interval, the highest metallicity that is consistent with the observed environmental density of SLSN hosts, and therefore the apparent upper limit of SLSN production is $12+\log($O/H$) = 8.12$. {This is approximately 0.2 dex lower than the apprent threshold reported in \citet{schulze2018}, $12+\log($O/H$) \sim 8.3$, and 0.3 dex lower than the threshold reported in \citet{Chen2017}, $12+\log($O/H$) \sim 8.4$.} 

In addition, we find that only about 26 per cent of BLM galaxies in groups reside in haloes of $M_h \geq 10^{13}\ \msun$. This supports the idea that the typical galaxies that host SLSNe will rarely be found in large groups or clusters.

The distributions in Figure \ref{fig:metallicity-ssfr} provide evidence that low-metallicity is the constraining requirement for SLSN production, rather than high-SFR. Focusing on the right panel, we see that the distribution of BLM-g in sSFR is more evenly distributed compared to metallicity in the left panel. In particular, noting SLSN host galaxies require sSFR$\gtrsim 10^{-9}\ $yr$^{-1}$ \citep{schulze2018}, there is actually an excess of BLM-g galaxies at this sSFR. This means that if a high-sSFR was the predominant factor for SLSN production compared to low-metallicity, we would see many more SLSN host galaxies in denser environments: $\approx 24$ out of the 33 SLSN events in this sample. Since this is not the case, our analysis suggests that low-metallicity is likely the more important requirement.

\section{Conclusions}\label{sec:conc}

Using SLSN sources from the ZTF BTS, we calculate whether their host galaxies are found in high-density or low-density large-scale environments. We use bright, spectroscopically redshifted galaxies with absolute magnitude $M_i<-21.5$ as tracers of density, and classify a source as in a high-density environment if there is 2 or more bright galaxies within 2 Mpc. We test our findings by comparing the SLSN sample with a sample of Type Ia supernovae, whose host galaxies span a wider range of galaxy colour and absolute magnitude. We also create another sample, derived from SDSS photometrically-redshifted galaxies, matched to the SLSN host galaxies in galaxy colour, absolute magnitude, and redshift. Our main findings may be summarised as follows:

\begin{itemize}
    \item SLSN host galaxies are almost always located in low-density environments, with all but 2 (of 33) host galaxies having no bright galaxies within 2 Mpc. The photometrically selected matched sample (in luminosity, colour, and redshift) shows similar results. This is in contrast to Type Ia supernova host galaxies, {8.5 per cent of which have 2 or more bright neighbours} This is consistent with the fraction of SNe Ia found in clusters compared to field galaxies, according to \citet{Larison2023}. In any relevant magnitude bin, the maximum high-density fraction is $<$ 10 per cent for the SLSN and matched samples, while for the SN-Ia sample the minimum is $>$ 20 per cent.
    \item In analysis of the Illustris-TNG simulations, over 70 per cent of blue, low-mass galaxies are found in the field, rather than in groups. These galaxies have low metallicities and high star-formation rates, which are typical of SLSN host galaxies. 
    \item Crucially, we find that in order to quantitatively match the rate of SLSN host galaxies found in dense environments, an additional condition on the metallicity of the host in the simulation is required. Taking the uncertainties into account, SLSN production is only favoured in galaxies with $12+\log($O/H$) = 8.12$. In contrast, selecting only high-sSFR galaxies would lead to an over-representation of high-density hosts in the simulation.

\end{itemize}

The fact that simulations suggest that galaxies capable of producing SLSNe and having high-sSFR would preferentially be found in groups, but such galaxies with low-metallicity would preferentially be found in the field, suggests that the metallicity of a galaxy is the more important factor in the occurrence of these explosive transients. However, this result will be strengthened with more observations of SLSNe and their host galaxies, which will be achieved over the coming years when new high-cadence surveys come online such as LSST at the Vera Rubin Observatory.

\section*{Acknowledgements}
The authors thank the anonymous reviewer for their helpful and thoughtful comments on this paper. CC acknowledges support from the School of Physics and Astronomy at the University of Birmingham. SLM acknowledges support from STFC grant ST/S000305/1 and UK Space Agency Grants No.~ST/Y000692/1 and ST/X002071/1. MN is supported by the European Research Council (ERC) under the European Union’s Horizon 2020 research and innovation programme (grant agreement No.~948381).

Funding for the Sloan Digital Sky 
Survey IV has been provided by the 
Alfred P. Sloan Foundation, the U.S. 
Department of Energy Office of 
Science, and the Participating 
Institutions. 

SDSS-IV acknowledges support and 
resources from the Center for High 
Performance Computing  at the 
University of Utah. The SDSS 
website is www.sdss.org.

SDSS-IV is managed by the 
Astrophysical Research Consortium 
for the Participating Institutions 
of the SDSS Collaboration including 
the Brazilian Participation Group, 
the Carnegie Institution for Science, 
Carnegie Mellon University, Center for 
Astrophysics | Harvard \& 
Smithsonian, the Chilean Participation 
Group, the French Participation Group, 
Instituto de Astrof\'isica de 
Canarias, The Johns Hopkins 
University, Kavli Institute for the 
Physics and Mathematics of the 
Universe (IPMU) / University of 
Tokyo, the Korean Participation Group, 
Lawrence Berkeley National Laboratory, 
Leibniz Institut f\"ur Astrophysik 
Potsdam (AIP),  Max-Planck-Institut 
f\"ur Astronomie (MPIA Heidelberg), 
Max-Planck-Institut f\"ur 
Astrophysik (MPA Garching), 
Max-Planck-Institut f\"ur 
Extraterrestrische Physik (MPE), 
National Astronomical Observatories of 
China, New Mexico State University, 
New York University, University of 
Notre Dame, Observat\'ario 
Nacional / MCTI, The Ohio State 
University, Pennsylvania State 
University, Shanghai 
Astronomical Observatory, United 
Kingdom Participation Group, 
Universidad Nacional Aut\'onoma 
de M\'exico, University of Arizona, 
University of Colorado Boulder, 
University of Oxford, University of 
Portsmouth, University of Utah, 
University of Virginia, University 
of Washington, University of 
Wisconsin, Vanderbilt University, 
and Yale University.

Based on observations obtained with the Samuel Oschin Telescope 48-inch and the 60-inch Telescope at the Palomar
Observatory as part of the Zwicky Transient Facility project. ZTF is supported by the National Science Foundation under Grants
No. AST-1440341 and AST-2034437 and a collaboration including current partners Caltech, IPAC, the Weizmann Institute for
Science, the Oskar Klein Center at Stockholm University, the University of Maryland, Deutsches Elektronen-Synchrotron and
Humboldt University, the TANGO Consortium of Taiwan, the University of Wisconsin at Milwaukee, Trinity College Dublin,
Lawrence Livermore National Laboratories, IN2P3, University of Warwick, Ruhr University Bochum, Northwestern University and
former partners the University of Washington, Los Alamos National Laboratories, and Lawrence Berkeley National Laboratories.
Operations are conducted by COO, IPAC, and UW.

The IllustrisTNG simulations were undertaken with compute time awarded by the Gauss Centre for Supercomputing (GCS) under GCS Large-Scale Projects GCS-ILLU and GCS-DWAR on the GCS share of the supercomputer Hazel Hen at the High Performance Computing Center Stuttgart (HLRS), as well as on the machines of the Max Planck Computing and Data Facility (MPCDF) in Garching, Germany.

\section*{Data Availability}

The data used for this analysis is available at the links in the text above where the data is described.


\bibliographystyle{mnras}
\bibliography{main} 



\appendix
\section{Appendix}
Table \ref{tab: slsnelist} lists all 55 SLSNe events that were available on the BTS when analysis began, the groups that classified them as SLSNe, and the relevant citation.

\clearpage
\onecolumn
\begin{center}
\begin{longtable}{lcr}
\caption{Table of ZTF IDs of the transient events that were obtained on the BTS, the group that classified them as SLSNe, and the relevant citation.}\label{tab: slsnelist}\\

\multicolumn{1}{c}{{ZTFID}} & \multicolumn{1}{c}{{Classifying group}} & \multicolumn{1}{c}{{Citation}} \\ \hline \hline
\endfirsthead

\multicolumn{3}{c}%
{{\bfseries \tablename\ \thetable{} -- continued from previous page}} \\
\multicolumn{1}{c}{{ZTFID}} & \multicolumn{1}{c}{{Classifying group}} & \multicolumn{1}{c}{{Citation}} \\ \hline \hline
\endhead

\hline \multicolumn{3}{r}{{Continued on next page}} \\ 
\endfoot

\hline \hline
\endlastfoot

\textbf{ZTF18aavrmcg} &NUTS &\citet{2018TNSCR.673....1D} \\
\textbf{ZTF18aapgrxo} &ZTF &\citet{2018TNSCR.815....1F} \\
\textbf{ZTF18abmasep} &ZTF &\citet{2018TNSCR1232....1F} \\
\textbf{ZTF18ablwafp} &ePESSTO &\citet{2018TNSCR1396....1G} \\
\textbf{ZTF18abshezu} &ZTF &\citet{2018TNSCR1411....1F} \\
\textbf{ZTF18abvgjyl} &ZTF &\citet{2018TNSCR1416....1F} \\
\textbf{ZTF18acapyww} &ZTF &\citet{2018TNSCR1870....1F} \\
\textbf{ZTF18acenqto} &ZTF &\citet{2018TNSCR1877....1F} \\
\textbf{ZTF18achdidy} &ZTF &\citet{2019TNSCR..32....1F} \\
\textbf{ZTF18acxgqxq} &ZTF &\citet{2019TNSCR..32....1F} \\
\textbf{ZTF18acyxnyw} &ZTF &\citet{2019TNSCR.188....1F} \\
\textbf{ZTF19aaknqmp} &ZTF &\citet{2019TNSCR.598....1F} \\
\textbf{ZTF19aanesgt} &ZTF &\citet{2019TNSCR.636....1F} \\
\textbf{ZTF19aacxrab} &ZTF &\citet{2019TNSCR.747....1F} \\
\textbf{ZTF19aavouyw} &None &\citet{2019TNSCR.938....1C} \\
\textbf{ZTF19aarphwc} &ZTF &\citet{2019TNSCR.952....1F} \\
\textbf{ZTF19aawfbtg} &TCD &\citet{2019TNSCR1598....1P} \\
\textbf{ZTF19abpbopt} &ZTF &\citet{2019TNSCR1646....1P} \\
\textbf{ZTF19abnacvf} &ZTF &\citet{2019TNSCR1712....1P} \\
\textbf{ZTF19abfvnns} &ZTF &\citet{2019TNSCR1774....1F} \\
\textbf{ZTF18aajqcue} &ZTF &\citet{2019TNSCR1838....1F} \\
\textbf{ZTF19abxgmzr} &ZTF &\citet{2019TNSCR1923....1F} \\
\textbf{ZTF19acfwynw} &C-SNAILS &\citet{2019TNSCR2271....1N} \\
\textbf{ZTF19acgjpgh} &TCD &\citet{2019TNSCR2339....1P} \\
\textbf{ZTF20aahbfmf} &SGLF &\citet{2020TNSCR.500....1P} \\
\textbf{ZTF20aaifybu} &ZTF &\citet{2020TNSCR1736....1Y} \\
\textbf{ZTF19acbonaa} &ZTF &\citet{2020TNSCR1737....1Y} \\
\textbf{ZTF20aayprqz} &ZTF &\citet{2020TNSCR1749....1P} \\
\textbf{ZTF20aattyuz} &ZTF &\citet{2020TNSCR1756....1D} \\
\textbf{ZTF20abobpcb} &SGLF &\citet{2020TNSCR2456....1P} \\
\textbf{ZTF20abpuwxl} &None &\citet{2020TNSCR2902....1T} \\
\textbf{ZTF20abzumlr} &ZTF &\citet{2020TNSCR3125....1P} \\
\textbf{ZTF20acphdcg} &ePESSTO+ &\citet{2020TNSCR3486....1I} \\
\textbf{ZTF20acpyldh} &FLEET &\citet{2020TNSCR3871....1B} \\
\textbf{ZTF21aaarmti} &ePESSTO+ &\citet{2021TNSCR..86....1G} \\
\textbf{ZTF21aagpymw} &ePESSTO+ &\citet{2021TNSCR.338....1M} \\
\textbf{ZTF19ackjrru} &FLEET &\citet{2021TNSCR.565....1G} \\
\textbf{ZTF19acyjzbe} &FLEET &\citet{2021TNSCR.565....1G} \\
\textbf{ZTF19adaivcf} &FLEET &\citet{2021TNSCR.565....1G} \\
\textbf{ZTF21aappdnv} &Global SN Project &\citet{2021TNSCR1220....1G} \\
\textbf{ZTF21aakjkec} &ZTF &\citet{2021TNSCR1234....1D} \\
\textbf{ZTF21aaxwpyv} &TCD &\citet{2021TNSCR1365....1D} \\
\textbf{ZTF21aavdqgf} &None &\citet{2021TNSCR1614....1Y} \\
\textbf{ZTF21abaiono} &ZTF &\citet{2021TNSCR1649....1P} \\
\textbf{ZTF21abcpsjy} &REFITT &\citet{2021TNSCR2270....1W} \\
\textbf{ZTF21abbqeea} &SGLF &\citet{2021TNSCR2271....1P} \\
\textbf{ZTF21absyjff} &FLEET &\citet{2021TNSCR3270....1G} \\
\textbf{ZTF21accwovq} &ePESSTO+ &\citet{2021TNSCR3651....1G} \\
\textbf{ZTF21abrqria} &FLEET &\citet{2021TNSCR3662....1G} \\
\textbf{ZTF21aalkhot} &ePESSTO+ &\citet{2022TNSCR.583....1S} \\
\textbf{ZTF22aabimec} &ePESSTO+ &\citet{2022TNSCR.667....1S} \\
\textbf{ZTF22aadqgoa} &SGLF &\citet{2022TNSCR1175....1P} \\
\textbf{ZTF22aalzjdc} &UCSC &\citet{2022TNSCR1881....1D} \\
\textbf{ZTF22aarqrxf} &ePESSTO+ &\citet{2022TNSCR2393....1A} \\
\textbf{ZTF22abcvfgs} &ePESSTO+ &\citet{2022TNSCR2997....1H} \\
\end{longtable}
\end{center}

\bsp	
\label{lastpage}
\end{document}